\documentclass[reprint,secnumarabic,amssymb,amsmath,aps,prb,groupedaddress,frontmatterverbose,showpacs]{revtex4-1}
\usepackage[dvips]{graphicx}
\usepackage{bm}
\usepackage{color}
\begin{document}
\title {Octahedral tilting and emergence of ferrimagnetism in cobalt-ruthenium based double perovskites}
\author{Manjil Das, Prabir Dutta, Saurav Giri and Subham Majumdar$^*$}
\email{$^*$sspsm2@iacs.res.in}
\affiliation{School of Physical Sciences, Indian Association for the Cultivation of Science, \\2A \& B Raja S. C. Mullick Road, Jadavpur, Kolkata 700 032, INDIA}

%%%%%%%%%%%%%%%%%%%%%%%%%%%%%%%%%%%%%%%%%%%%%%%%%%%%%% Abstract %%%%%%%%%%%%%%%%%%%%%%%%%%%%%%%%%%%%%%%%%%%%%%%%%%%%%%%%%%%%%%%%
\begin{abstract}
Rare earth based cobalt-ruthenium double perovskites A$_2$CoRuO$_6$ (A = La, Pr, Nd and Sm)  were synthesized and investigated for their structural and  magnetic properties. All the  compounds crystallize in the monoclinic $P2_1/n$ structure with the indication of antisite disorder between Co and Ru sites. While, La compound is already reported to have an antiferromagnetic state below 27 K, the Pr, Nd and Sm systems are found to be ferrimagnetic below $T_c$ = 46, 55 and 78 K respectively. Field dependent magnetization data indicate prominent hysteresis loop below $T_c$ in the  samples containing magnetic rare-earth ions, however magnetization does not saturate even at the highest applied fields. Our structural analysis indicates strong distortion in the Co-O-Ru bond angle, as La$^{3+}$ is replaced by smaller rare-earth ions such as Pr$^{3+}$, Nd$^{3+}$ and Sm$^{3+}$. The observed ferrimagnetism is possibly associated with  the enhanced antiferromagnetic superexchange interaction in the Co-O-Ru pathway due to bond bending. The Pr, Nd and Sm samples also show small magnetocaloric effect with Nd sample showing highest value of magnitude $\sim$ 3 Jkg$^{-1}$K$^{-1}$ at 50 kOe. The change in entropy below 20 K is found to be positive in the Sm sample as compared to the negative value in the Nd counterpart.    
  
\end{abstract}

\maketitle
\section{Introduction}
%%%%%%%%%%%%%%%%%%%%%%%%%%%%%%%%%%%%%%%%%%%%%%%%%%%% Introduction %%%%%%%%%%%%%%%%%%%%%%%%%%%%%%%%%%%%%%%%%%%%%%%%%%%%%
Since the discovery of low field room temperature magneto-resistance in Sr$_2$FeMoO$_6$~\cite{kobayashi}, the double perovskites (A$_2$BB$^{\prime}$O$_6$) have been intensely studied~\cite{vasala}. These quaternary compounds can be synthesized with a varying combinations of cations at the A (alkaline earth or rare earth metals) and B/B$^{\prime}$ (3$d$, 4$d$ or 5$d$ transition metals) sites, which provides a scope to access diverse material properties within the  similar crystallographic environment. Elements with partially filled $d$ level at the B/B$^{\prime}$ can give rise to wealth of magnetic ground states including ferromagnetism, antiferromagnetism, ferrimagnetism as well as glassy magnetic phase. Double perovskites are also associated with intriguing electronic  properties~\cite{ijp} such as half-metallic behaviour~\cite{serrate}, tunneling magneto-resistance~\cite{tmr}, metal-insulator transition~\cite{m-i} and so on.
\par
In case of insulating A$_2$BB$^{\prime}$O$_6$ double perovskites, the magnetic interaction between B and B$^{\prime}$ ions is primarily B-O-B$^{\prime}$ superexchange type and Goodenough-Kanamori rule can predict the sign of the interaction~\cite{dass}. Ideal double perovskites have cubic symmetry, but the presence of cations with small ionic radii at the A site can distort the structure. Such distortion lowers the lattice symmetry to tetragonal or monoclinic, where the   BO$_6$/B$^{\prime}$O$_6$ octahedra get tilted through the bending of B-O-B$^{\prime}$ bond angle.  It has been found that for two fixed B and B$^{\prime}$, the magnetic ground state is very much sensitive to this bond angle. Doping at the A site can change the bond angle and henceforth the nature of the ordered magnetic state. For example, substitution of Ca at the Sr site of Sr$_2$CoOsO$_6$ drives the system from an antiferromagnetic (AFM) insulator to spin-glass (SG) and eventually to a ferrimagnetic (FI) state on full replacement of Sr by Ca~\cite{srcacooso6,srcafeoso6}. There is also report of drastic change in ferrimagnetic coercivity in (Ca,Ba)$_2$FeReO$_6$ under hydrostatic pressure due to the buckling of Fe-O-Re bond~\cite{a2fereo6}. It has been argued that the magnetic ground state in these insulating systems is an outcome of the competition between the interactions along the paths B-O-B$^{\prime}$ and B-O-B$^{\prime}$-O-B (and similarly, B$^{\prime}$-O-B-O-B$^{\prime}$). When the octahedral tilting is minimal, the B-O-B$^{\prime}$-O-B type interaction dominates, giving rise to strong AFM correlations within B sublattice~\cite{kanungo}. However, with increasing distortion, B-O-B$^{\prime}$ becomes stronger with the simultaneous weakening of B-O-B$^{\prime}$-O-B coupling, and a simple FI state emerges (provided the magnetic moments at B and B$^{\prime}$ ions are unequal) due to the strong AFM coupling between B and B$^{\prime}$ ions. The intermediate spin-glass state possibly arises from these competing interactions. 

\par
Recently, La$_2$CoRuO$_6$ compound has been shown to have an AFM ground state, which crystallizes in the distorted monoclinic structure~\cite{bos}. Interestingly, the isostructural Y$_2$CoRuO$_6$ shows ferrimagnetism, and turns into a spin-glass on La doping at the Y site~\cite{y2coruo6}. It is therefore worthwhile to study the magnetic states of A$_2$CoRuO$_6$ with different rare-earth atoms at the A site. It is well known that the ionic radius of rare-earth (in the 3+ state) diminishes with increasing atomic number, which can tune the lattice distortion. The main goal of the present work is to study the effect of lattice distortion and associated change in the magnetic properties with varying rare-earth ion at the A site. It is also worthwhile to note how the rare-earth moment is affecting the ground state magnetic properties. In a recent report, the magnetic properties of A$_2$CoMnO$_6$ compounds were found  to  get strongly affected by the variation of rare-earth ion at the A-site~\cite{tknath}.     

%%%%%%%%%%%%%%%%%%%%%%%%FIGURE 1%%%%%%%%%%%%%%%%%%%%%%%%
\begin{figure*}[t]
\begin{center}
\includegraphics[width = 14 cm]{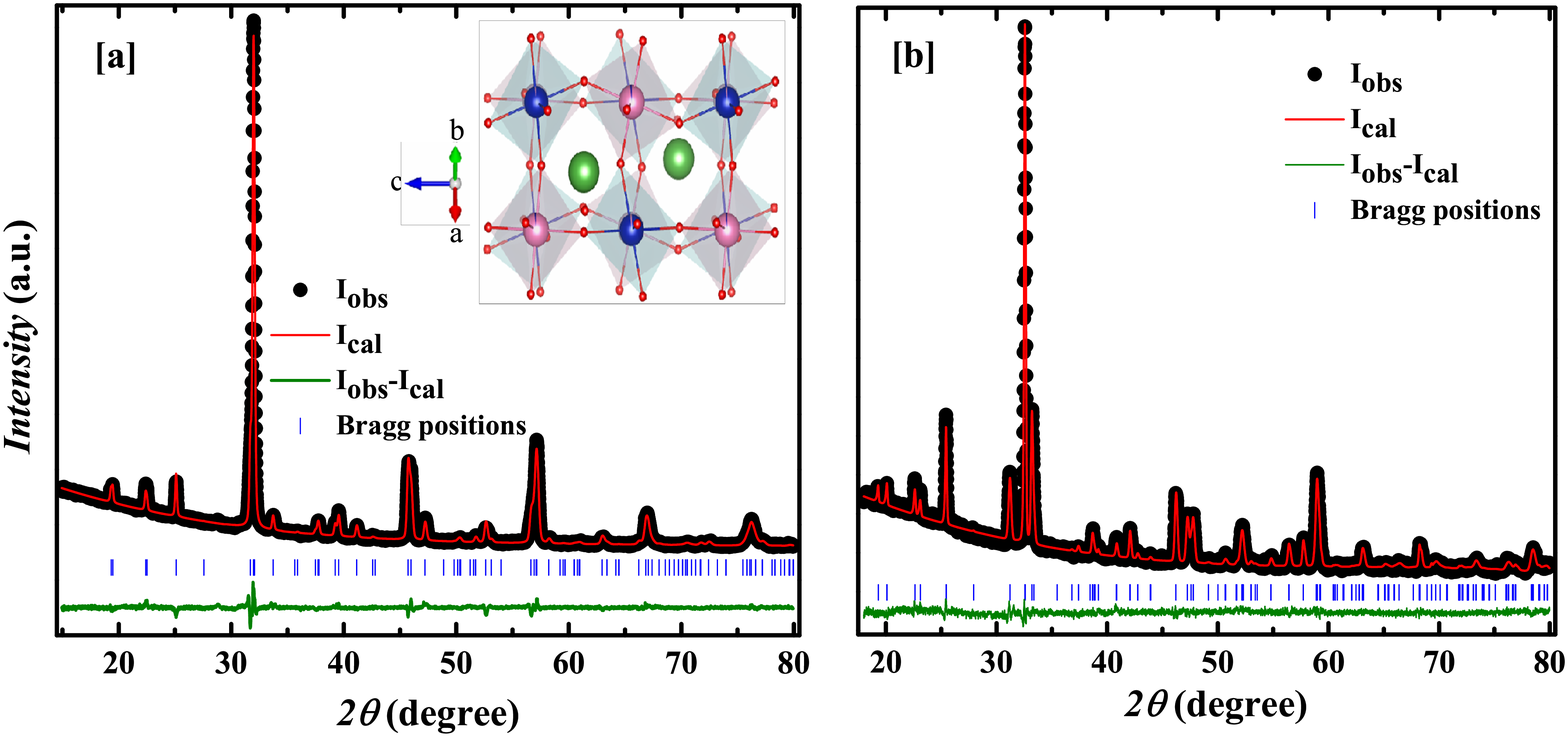}
\vskip 1.4cm 
\includegraphics[width = 10 cm]{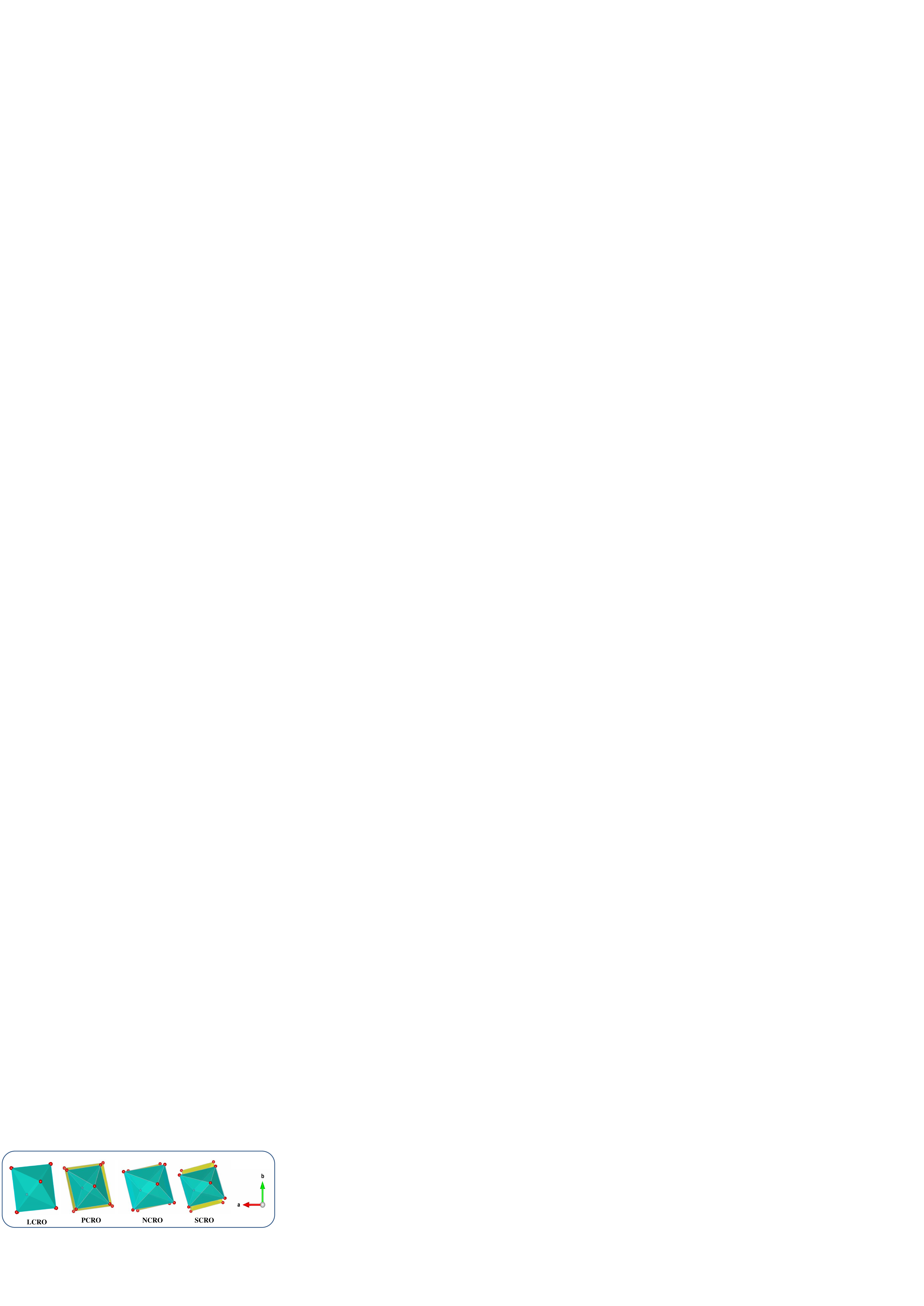}
\caption {(a) and (b) show powder x-ray  diffraction data of LCRO and SCRO respectively collected at room temperature. The  inset of (a) shows the crystal structure of the sample. Green, pink, blue and  spheres indicate La, Ru, Co and O atoms respectively. The bottom panel indicates the octahedral tilt of four compositions.}
\end{center}
\end{figure*}
%%%%%%%%%%%%%%%%%%%%%%%%%%%%%%%%%%%%%%%%%%%%%%%%%%%%%%%%

%%%%%%%%%%%%%%%%%%%%%%%%FIGURE 2%%%%%%%%%%%%%%%%%%%%%%%%
\begin{figure*}[t]
%\begin{center}
\includegraphics[width = 14 cm]{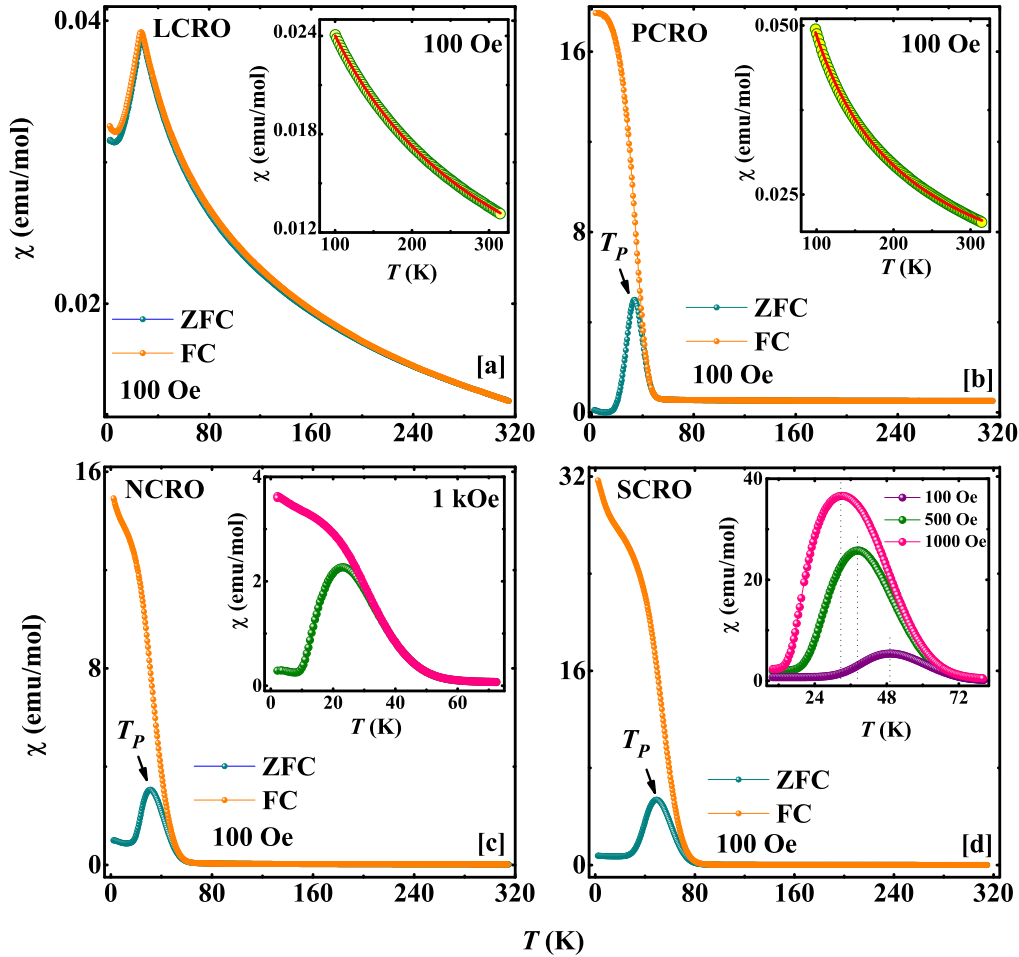}
\caption {(a)-(d) represent temperature variation of magnetic susceptibility for LCRO, PCRO, NCRO and SCRO respectively for an applied field of 100 Oe both in ZFC and FC protocols. The insets of (a) and (b) show the modified Curie-Weiss fit to the susceptibility data of respective samples. The inset of (c) shows the susceptibility of NCRO measured under 1 kOe of field. The inset in (d) shows an enlarged view of the ZFC susceptibility of SCRO  recorded at different applied fields.}
%\end{center}
\end{figure*}
%%%%%%%%%%%%%%%%%%%%%%%%%%%%%%%%%%%%%%%%%%%%%%%%%%%%%%%% 

\section{Experimental Details}
Single phase polycrystalline A$_2$CoRuO$_6$  samples (for A = La, Pr, Nd and Sm) were synthesized by solid state technique. Stiochiometric amounts of A$_2$O$_3$ (except Pr-sample, where Pr$_6$O$_{11}$ was used), Co$_3$O$_4$ and RuO$_2$ were well mixed in a agate morter pestle and calcined at 1073 K for 12 h and then sintered at 1473 K for 24 h in the pellet form with one intermediate grinding.  The powder X-ray  diffraction (PXRD) data were obtained in RIGAKU X-ray diffractometer with Cu-K$_{\alpha}$ radiation in the range 15$^{\circ}$ to 80$^{\circ}$. Magnetization ($M$) measurements were carried out  using a SQUID magnetometer (Quantum Design, MPMS-3) up to 70 kOe. High field magnetic measurements (up to 150 kOe) was performed on a vibrating sample magnetometer from Cryogenic Ltd., UK.  The temperature dependence of the electrical resistivity ($\rho$) was measured by  DC four-probe method in the temperature range between 50 K and 300 K.

\section{Sample Characterization}

All four compositions, La$_2$CoRuO$_6$ (LCRO), Pr$_2$CoRuO$_6$ (LCRO) Nd$_2$CoRuO$_6$ (NCRO) and Sm$_2$CoRuO$_6$ (SCRO),  crystallize in monoclinic  rock salt structure (space group $P2_1/n$)~\cite{anderson}. For double perovskite A$_2$BB$^{\prime}$O$_6$, one can define a Goldschmidt tolerance factor $t = \frac{r_A + r_O}{\sqrt{2}(r_B + r_O)}$, where $r_A$, and $r_O$ are the ionic radii of A and O respectively, while $r_B$ stands for the average radius of B and B$^{\prime}$. It has been found that if $t <$ 1, there can be distortion from the ideal cubic structure~\cite{vasala}.  For the present case, $r_{La}^{3+}$= 103.2 pm, $r_{Pr}^{3+}$= 99 pm $r_{Nd}^{3+}$= 98.3 pm, $r_{Sm}^{3+}$ = 95.8 pm, $r_{Co}^{2+}$  = 65(74.5) pm for low-spin (high-spin), $r_{Ru}^{4+}$ = 62 pm and $r_O^{2-}$ = 140 pm, which give $t$ in the range of  0.81-0.83 for four samples. Evidently for these compositions, $t$ is significantly lower than unity, and this explains the observed monoclinic symmetry rather than ideal cubic one. This lower symmetry is associated with the tilting of the (B,B$^{\prime}$)O$_6$ octahedra. In order to determine the crystallographic parameters of our samples, we have performed Reitveld refinement on the room temperature PXRD data using MAUD software package~\cite{maud}. The data converges well with monoclinic space group P2$_1$/n for all the samples along with antisite disorder between Co and Ru sites [Figs. 1 (a) and (b)]. Our calculations show that there are ~1-5\% antisite defects, {\it i.e.}, the fractional occupancy of B site consists of 99-95\% Co and 1-5\% Ru. The antisite defect is found to be large in Nd and Sm compounds and it is low in case of La and Pr  counterparts. The refined crystallographic parameters are depicted in Table 1, and they match well with the previous reported data~\cite{kawano}. We have also added the structural data of Y$_2$CoRuO$_6$ (YCRO) from reference~\cite{y2coruo6} for comparison. The crystal structure of LCRO, as obtained from the refinement of our PXRD data, is shown in the inset of fig. 1 (a). It is clearly evident that the (Co,Ru)O$_6$ octahedra are tilted. The average tilting angle is defined as $\langle \Psi \rangle = \frac{1}{2}[\pi-\langle \Phi \rangle]$, where $\langle \Phi \rangle$ is the average inter-octahedral Co-O-Ru angle. Clearly, the tilt angle increases systematically as we move from La to Sm, which is the effect of the gradual bending of Co-O-Ru bond (see Table 1).  It is interesting to note that all the crystallographic parameters vary systematically with the ionic radius for La, Pr, Nd and Sm samples. 

\begin{table*}
\centering
\begin{tabular}{|c|c|c|c|c|c|}
\hline

Parameters & LCRO & PCRO & NCRO & SCRO & YCRO \\ 
\hline
$r_{A}^{3+}$ (\AA)   &  1.032 & 0.990 &  0.983   & 0.958 & 0.900   \\
$a$ (\AA)      &  5.575(2) & 5.497(1)   & 5.440(5)   & 5.390(2) & 5.266   \\
$b$ (\AA)     &  5.638(7) & 5.689(7)  & 5.717(7)   &5.722(8) & 5.711   \\
$c$ (\AA)      &  7.886(2) & 7.802(4)   & 7.739(2)   &7.685(7) & 7.558   \\
$\beta$($^{\circ}$)  & 90.01(1) & 89.88(7)    & 89.99(8)   &89.93(2) & 90.03  \\
$\langle \angle${Co-O-Ru} $\rangle$ ($^{\circ}$) & 152.4 & 150.9 & 143.4 & 139.5 & 141.7\\
$\langle \Psi \rangle$($^{\circ}$) & 13.8 & 14.6 & 18.3 & 20.3 & 19.7 \\
\hline
Mag. state &  AFM & FI   & FI   & FI & FI \\
Tran. temp.& $T_N$ = 27 K & $T_c$ = 46 K   &$T_c$ = 55 K &  $T_c$ = 78 K & $T_c$ = 82 K \\
$H_{coer}$ & -- & 9 kOe (2 K) & 8 kOe (2 K) & 22 kOe (2 K) & 22.5 kOe (5 K) \\
 
\hline\hline

\end{tabular}
\caption{Crystallographic lattice parameters ($a$, $b$, $c$, and $\beta$), average Co-O-Ru bond angle, average octahedral tilt angle ($\Psi$), magnetic state, magnetic transition temperatures and coercivity are depicted for A$_2$CoRuO$_6$ (A = La, Pr, Nd, Sm and Y). The parameters for Y compound are obtained from reference~\cite{y2coruo6}. The ionic radii of rare-earth ions (in the 3+ state with coordination number VI) at the A site are also shown~\cite{ionicradii}.}
\end{table*}

\section{Results}
\subsection{Magnetic studies}

Figs. 2 (a)-(d) depict the temperature ($T$) dependence of susceptibility ($\chi = M/H$) measured under different values of $H$ values for all the four samples, where both zero-field-cooled (ZFC) and field-cooled (FC) measurements were performed. LCRO [fig.2 (a)] shows well defined peak at the Ne\'el temperature $T_N$ = 27 K, indicating an AFM ground state and it matches well with the previous report~\cite{dass,bos,yoshii1}. The $T$ variation of  susceptibility ($\chi = M/H$) of LCRO in the paramagnetic (PM) state cannot be fitted with a simple Curie-Weiss law. However, the $\chi(T)$ data above 100 K can be fitted well with a modified Curie-Weiss law, $\chi^{CW}(T) = C/(T-\theta) + \chi_0$, where an additional $T$ independent term ($\chi_0$) is included. Here $C$ is the Curie constant and $\theta$ is the Curie-Weiss temperature. The effective PM moment, $\mu_{eff}$, obtained from Curie-Weiss fitting, is found to be 6.63 $\mu_B$/f.u. The value of $\theta$ is $-$140 K, signifying strong AFM correlations. The FC and ZFC data show weak divergence below $T_N$. We also observed an upward rise in the $\chi(T)$ data below 6 K. The observed value of $\mu_{eff}$ is higher than expected for a Co$^{2+}$-high-spin and Ru$^{4+}$-low-spin states~\cite{dass}, which was attributed to extended 4$d$-orbitals of Ru~\cite{yoshii1}. 

\par 
For PCRO, NCRO and SCRO, the $\chi(T)$ data are drastically different from that of LCRO [figs. 2 (b), (c) and (d) respectively], and it is quite eventful. $\chi$ shows a sharp rise below $T_c$ = 46, 55 and 78  K for these magnetic rare-earth containing samples respectively.  The FC and ZFC susceptibilities show strong irreversibility below $T_c$. The divergence  exists even in measurement at $H$ = 1 kOe [see inset of fig. 2 (c)], however the extend of divergence reduces. The point of bifurcation also moves to lower $T$ with increasing $H$. The ZFC data show a well defined peak at a temperature $T_P$, which lies  below $T_c$.  Notably, we observe a change in the value of $T_P$ with increasing $H$. The $H$ dependence of $T_P$ is particularly significant for SCRO, where we observe a shift of $T_p$ by 16 K when $H$ is changed from 100 Oe to 1 kOe [see inset of fig. 2 (c)]. Similar shift in the observed peak in ferrimagnetic Nd$_2$CoMnO$_6$ was also reported previously, which was attributed to the presence of ferromagnetic (FM) and AFM clusters~\cite{ncmo}. The FC susceptibility, on the other hand, rises monotonically for all the samples with decreasing $T$. 

\par
The susceptibility data of PCRO and NCRO can be well fitted with $\chi^{CW}(T)$ above 100 K, which gives the values of $\mu_{eff}$ to be 6.59 and 6.47 $\mu_B$ respectively. The value of  $\theta$ is -25 (-22) K for Pr(Nd) sample. These values of  $\mu_{eff}$ are slightly lower than the  expected value of 6.97 (7.01) $\mu_B$  with Pr$^{3+}$ (Nd$^{3+}$), Co$^{2+}$ (high-spin) and Ru$^{4+}$ (low-spin) states. This mismatch can be caused by  the presence of antisite disorder. Such disorder is expected to cause a decrease in the magnetic moment, and empirically  $M_{actual} = M_{obs}/(1-2\mathcal{D})$~\cite{feng} where $\mathcal{D}$ is the degree of  disorder between Co and Ru atoms. For example, in case of NCRO  with  $\mathcal{D}$ = 0.05 and $M_{obs}$ = 6.47 $\mu_B$,  $M_{actual}$ is found to be 7.18  $\mu_B$,  which is very close to the theoretically predicted value. 

\par
For  SCRO, we failed to achieve good fit using $\chi^{CW}(T)$ in the temperature range 100-315 K. The  separation between  ground ($J$ = 5/2) and first excited ($J$ = 7/2) multiplets in Sm$^{3+}$  is small, and their mixing can be responsible for the observed non-Curie-Weiss behaviour~\cite{wijn}. 

\begin{figure*}[t]
\begin{center}
\includegraphics[width = 14 cm]{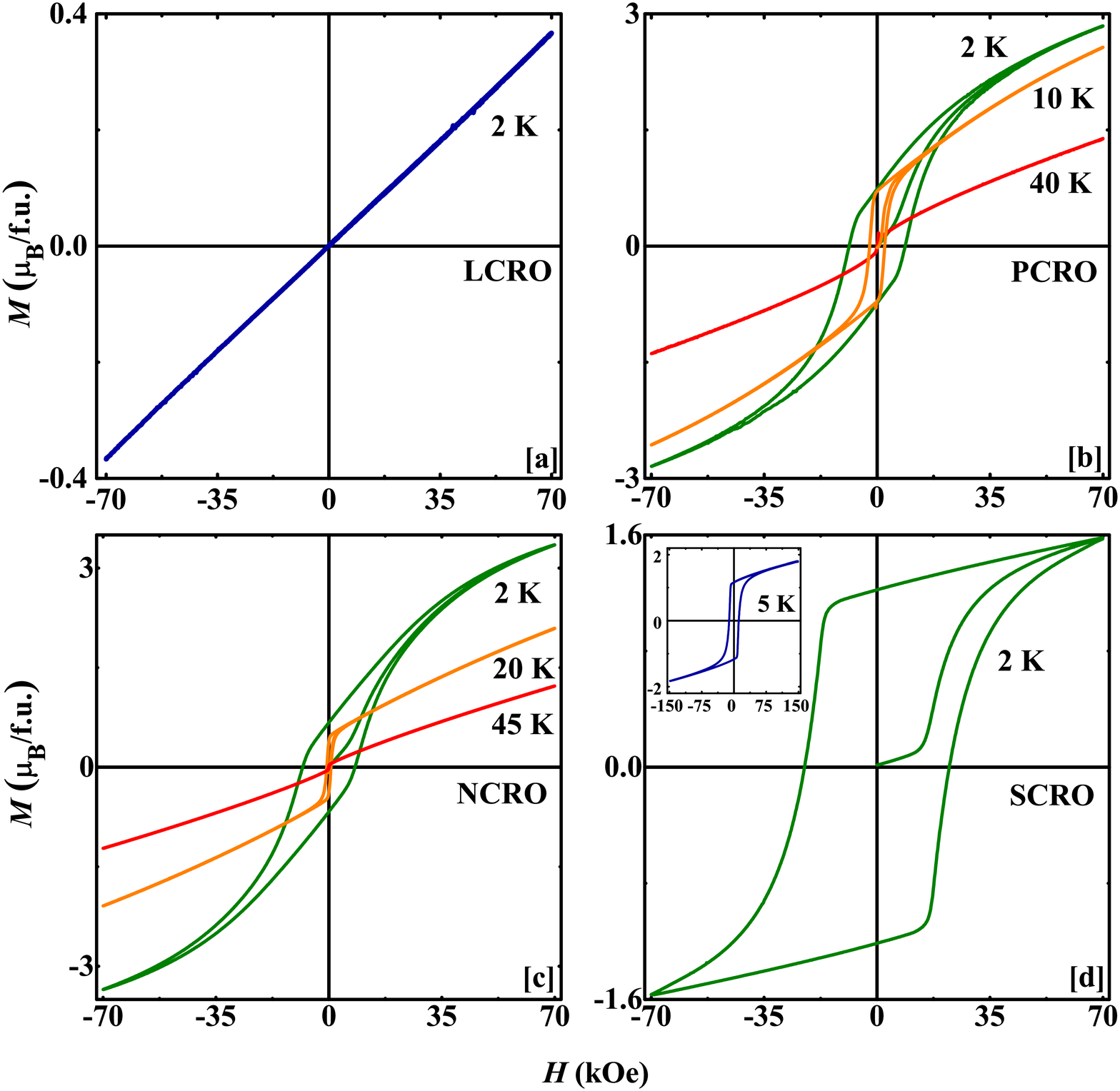}
\caption {(a) to (d) show isothermal magnetization data up to field 70 kOe at different temperatures for LCRO, PCRO, NCRO and SCRO  respectively. The inset of (d) shows the $M-H$ curve for SCRO at 5 K for maximum field of 150 kOe.}
\end{center}
\end{figure*}
%%%%%%%%%%%%%%%%%%%%%%%%%%%%%%%%%%%%%%%%%%%%%%%%%%%%%%%%
%%%%%%%%%%%%%%%%%%%%%%%%FIGURE 4%%%%%%%%%%%%%%%%%%%%%%%%
\begin{figure*}[t]
\begin{center}
\includegraphics[width = 14 cm]{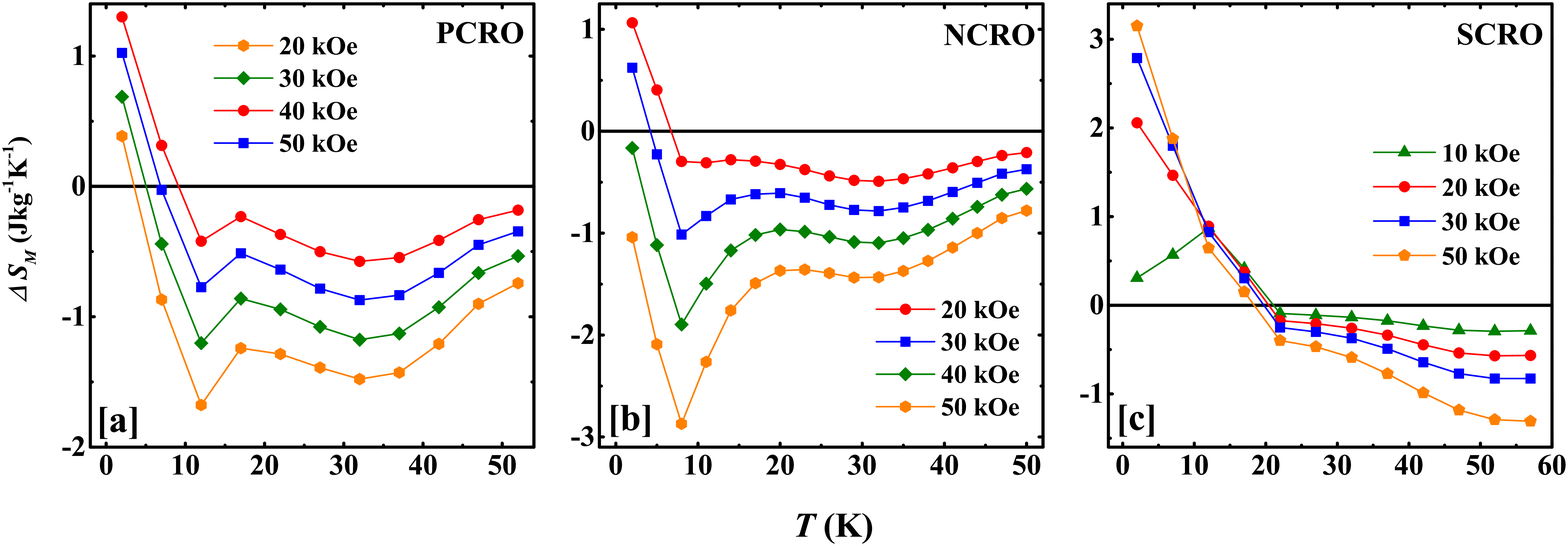}
\caption {(a) to (c) respectively show the $T$ variation of $\Delta S_M$ of PCRO, NCRO and SCRO at different magnetic fields.}
\end{center}
\end{figure*}
%%%%%%%%%%%%%%%%%%%%%%%%%%%%%%%%%%%%%%%%%%%%%%%%%%%%%%%%
\par 
The isothermal $M$ versus $H$ data are shown in figs. 3 (a) to (d). For LCRO, a linear $M-H$ curve is obtained at 2 K [see fig. 3 (a)], indicating AFM state. On the other hand, Pr, Nd and Sm compounds show significant hysteresis with large coercive field ($H_{coer}$). Non-zero $H_{coer}$ is observed for  these FI systems just below $T_c$, and it increases with decreasing $T$. The values of $H_{coer}$ are found to be 9, 8  and 22 kOe at 2 K for PCRO, NCRO and SCRO respectively. However, $M$ does not saturate even at 70 kOe of field for none the samples. We have also recorded high field magnetization for SCRO, as shown in the  inset of fig. 3 (d). The $M-H$ curve at 5 K does not fully saturate even at 150 kOe of applied field. $M$ attains a value  close to 2 $\mu_B$ at the highest  $H$. The value of $H_{coer}$ for SCRO at 5 K is found to be 10.6 kOe, which  is  smaller than the value of $H_{coer}$ reported for Y$_2$CoRuO$_6$ ($\sim$ 22.5 kOe) at the same $T$.     

\par
Considering non-zero coercivity and the presence of antisite disorder, we have recorded  $M-H$ hysteresis loop at 2 K after the sample being field-cooled from room temperature. In case of inhomogeneous magnetic systems, a shift in the hysteresis loop along the field axis  may be  observed due to the interfacial coupling of two magnetic phases, and it is referred as exchange bias effect~\cite{giri}. Many double perovskites show exchange bias effect due to the presence of antisite disorder~\cite{eb}. However, we failed to observe such exchange bias in NCRO and SCRO samples, which possibly rule out the existence of large magnetic inhomogeneity in the system.    

\par
In order to investigate the effect of external field on the magnetic state, we have measured magneto-caloric effect (MCE) of the samples in terms of  entropy-change ($\Delta{S_M}$) by  $H$. In the recent past, MCE has emerged out to be an important technique for green refrigeration~\cite{balli}. In the present work, we have obtained MCE from our isothermal magnetization data recorded at different constant temperatures. From the theory of thermodynamics, $$\Delta{S_M}(0\rightarrow H_0) = \int^{H_0}_{0}\left(\frac{\partial{M}}{\partial{T}}\right )_HdH,$$ where $\Delta{S_M}(0\rightarrow H_0)$ denotes the entropy change for the change in $H$ from 0 to $H_0$~\cite{mce}. Figs. 4 (a) to  (c) show $\Delta S_M (T,H_0)$ versus $T$ plot at different values of $H_0$ for PCRO, NCRO and SCRO samples respectively. 
\par
The magnitude of MCE is found to be low for all three samples. For the Pr and Nd samples, $\Delta{S_M}(T)$ is mostly negative with its magnitude peaking around 12 and 8 K (peak magnitude: 1.7 and 2.9 Jkg$^{-1}$K$^{-1}$ at $H_0$ = 50 kOe) respectively. A broad feature is also observed in $\Delta S_M (T,H_0)$ data around 35 K. On the other hand, SCRO shows a contrasting behaviour as far as the MCE is concerned. $\Delta{S_M}(T)$ for  SCRO is positive below 20 K, and increases with decreasing temperature (at least for $H_0 >$ 10 kOe). $\Delta S_M$ attains a value of 3.1 Jkg$^{-1}$K$^{-1}$ for $H_0$ = 50 kOe at 2 K. 

\par
As already discussed, double perovskite systems can show glassy magnetic state~\cite{anil}. An intermediate SG state is observed when the  AFM state is transformed into an FI state by suitable A-site doping. In order to investigate the possibility of a glassy state, particularly those which lie across the AFM-FI boundary, field-cooled-field-stop memory measurement were performed on Pr and Nd samples~\cite{memory1,memory2}. In this protocol, the samples were cooled in 100 Oe of field down to 2 K  with intermediate stops at several temperatures below $T_c$. Subsequently, the samples were heated back in 100 Oe and dc magnetization was measured. We do not observe any anomaly at the stopping temperatures during heating, which rules out the possibility of any glassy (spin glass or cluster glass) or super paramagnetic state in PCRO and NCRO samples.

%%%%%%%%%%%%%%%%%%%%%%%%FIGURE 5%%%%%%%%%%%%%%%%%%%%%%%%
\begin{figure*}[t]
\begin{center}
\includegraphics[width = 12 cm]{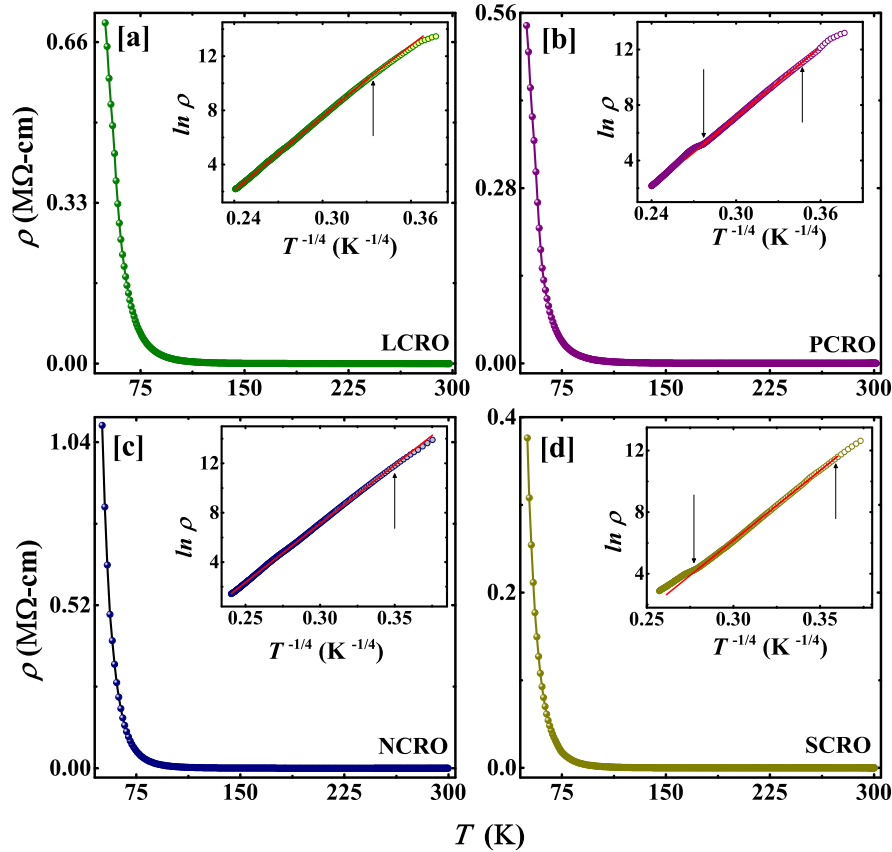}
\caption {(a) to (d) respectively show resistivity data as a function of temperature for LCRO, PCRO, NCRO and SCRO. The insets show the VRH-type fittings to the data.}
\end{center}
\end{figure*}
%%%%%%%%%%%%%%%%%%%%%%%%%%%%%%%%%%%%%%%%%%%%%%%%%%%%%%%%

\subsection{Electrical transport}
Like many  other A$_2$BB$^{\prime}$O$_6$ compounds, LCRO, PCRO, NCRO and SCRO show semiconducting behaviour as evident from the transport data depicted in figs. 5 (a) to (d) respectively. The values of $\rho$ at room temperature ($\sim$ 300 K) are found to be 8.86, 8.81,  4.16 and  6.68 $\Omega$-cm for La, Pr, Nd and Sm compounds. Our analysis on the $\rho(T)$ data indicates that all three compositions show Mott Variable Range (VRH) hopping conduction, where $\rho(T) \sim \exp\left[\left (\frac{T_0}{T} \right)^{\frac{1}{4}}\right ]$. This is also quite common among disordered double perovskite such as Sr$_2$MnRuO$_6$~\cite{woodward} or Sr$_2$CoSbO$_6$~\cite{martin}. The VRH type conduction is quite clearly visible from the $\log{\rho}$ versus $T^{-1/4}$ plots  in the respective insets of figs. 5 (a), (b) and (c). While for La and Nd compounds, the VRH nature is present almost over the full range of temperature (it deviates only below 65 K), Pr and Sm compounds show VRH conduction only in the range 160 to 60 K. The values of the parameter $T_0$ associated with the VRH conduction are found to be 6.9 $\times10^7$, 5.1$\times10^7$,   8.8 $\times10^7$ and 6.8 $\times10^7$ K for La, Pr, Nd and Sm compounds respectively.    

\section{Discussions}
It turns out that the magnetic properties of samples containing magnetic rare-earth  are drastically different from that of LCRO. Naively, one can relate  the FI state with the magnetic moment of A site. However, FI state is also observed in Y$_2$CoRuO$_6$, where A site contains nonmagnetic Y$^{3+}$ ions. In order to address the issue, let us first discuss various salient observations made on the studied samples.
\begin{enumerate}
 \item {The Co-O-Ru bond distortion (and consequently, the octahedral tilt) is found to increase  as we move from LCRO to heavier rare-earth containing compounds. This is due to the reduction of ionic radius at the A-site (lanthanide contraction), as we proceed from La to Sm. It has been  already mentioned that the smaller radius of A-element leads to larger B-O-B$^{\prime}$ bond distortion~\cite{vasala}.}
 
 \item{LCRO with relatively smaller bond distortion shows AFM ground state. On the other hand, PCRO, NCRO and SCRO show large increase in $M$ below the magnetic transition at $T_c$. Isothermal magnetization curves show large hysteresis with the presence of significant remanent magnetization. Isostructural Y$_2$CoRuO$_6$ shows similar magnetic behaviour and the Co-O-Ru superexchange interaction is found to be AFM in nature leading to a FI ground state (Co and Ru sublattices have different moment values)~\cite{y2coruo6}. In analogy with the Y compound, we can conclude that the ground states of PCRO, NCRO and SCRO are ferrimagnetic. Similar to Y compound, FI state in PCRO and subsequent compounds emerges possibly due to the Co-O-Ru bond distortion, rather than A-site magnetic moment formation.}
 
 \item{All four studied compounds are found to be semiconducting with reasonably high resistivity ($\sim$ M$\Omega$-cm)  around the magnetic anomalies. Therefore, the magnetic interaction is likely to be mediated by the superexchange, rather than the double-exchange mechanism.} 
 
 \item{The coercivity associated with $M-H$ hysteresis loop is found to be much higher in case of SCRO, however it is lower than the value reported for Y$_2$CoRuO$_6$. It appears that the coercivity is not directly connected to the fact whether A site contains a magnetic (such as Pr, Nd or Sm) or nonmagnetic (here Y) ion.}
 
 \item{$\Delta S_M$ versus $T$ plots for PCRO, NCRO and SCRO  show further anomaly at around 8-12 K. It is difficult to guess the origin of such anomaly. However, considering the presence of  rare-earth at the A-site, it may signify the low-$T$ ordering of the magnetic rare-earth ions.}

\end{enumerate}

\par
As already mentioned, a transition from AFM to FI via a glassy magnetic state has been observed in several double perovskites with the bending of the B-O-B$^{\prime}$ bond~\cite{srcacooso6,srcafeoso6,a2fereo6,y2coruo6}, where the lattice distortion was created by systematic doping at the A site or by applying hydrostatic pressure. In the present case, we found similar effect when one rare-earth ion is replaced by another one with smaller ionic radius. In analogy with the idea mooted in case of Sr$_{2-x}$Ca$_x$FeOsO$_6$ and  Sr$_{2-x}$Ca$_x$CoOsO$_6$~\cite{srcacooso6,srcafeoso6}, the AFM state in LCRO is due to the strong AFM correlation along the long bonds Co-O-Ru-O-Co and Ru-O-Co-O-Ru when the Co-O-Ru bond distortion is low. Replacement of La by Pr  initiates strong bending  in Co-O-Ru bond (see Table 1), which possibly  strengthen the  Co-O-Ru superexchange over the magnetic interaction on longer Co-O-Ru-O-Co and Ru-O-Co-O-Ru pathways leading to FI state. The bending further enhances in Sm compound, and a significantly large coercive field is observed.

\par 
The above argument is also supported by the drastic change in the values of  paramagnetic Curie temperature $\theta$ in Pr and Nd compounds (-25 and -22 K respectively) as compared to the antiferromagnetically ordered La counterpart. This may be an indication of the weakening of the exchange interaction along longer Co-O-Ru-O-Co and Ru-O-Co-O-Ru pathways. However, $\theta$ still remains negative due to the presence of  Co-O-Ru AFM interaction.   

\par
In case of SrCaCoOsO$_6$ and La$_{2-x}$Y$_x$CoRuO$_6$ ($x$ $\approx$ 0.25 to 1.5) SG states are observed~\cite{srcacooso6,y2coruo6}, when Ca(Y)  is doped at the Sr(La) site, which has been assigned due to the  frustration between long (B-O-B$^{\prime}$-O-B) and short(B-O-B$^{\prime}$) exchange paths. However, we do not see any glassy magnetic state in neither of the  Pr and Nd compounds. For La$_{1.25}$Y$_{0.75}$CoRuO$_6$ with tilt angle 15.5$^{\circ}$, a prominent frequency dispersion is observed in the ac susceptibility data. On the other hand PCRO with lower tilt angle (14.6$^{\circ}$) shows ordered FI state. Possibly, the emergence of glassy state in doped samples is also connected with the doping induced disorder. It is to be noted that  La$_{1.25}$Y$_{0.75}$CoRuO$_6$ sample has huge antisite disorder of 20\%, as compared to 1\% antisite disorder in PCRO.  

\par
It is now pertinent to address the role of 4$f$ moment from the rare-earth present at the A site. In case of isostructural  Er$_2$CoMnO$_6$, rare-earth moment orders at a relatively lower $T$ than the Co-Mn ordering temperature~\cite{ecmo}. From our magnetization data, it is hard to identify the ordering of rare-earth moment. The $\Delta S_M$ versus $T$ data depicted in fig.4 show peak like anomalies between  8  and 12 K in PCRO, NCRO and SCRO , and they can be probable ordering points of rare-earth moments. In order to shed more light on this issue, we have compared the moments of A$_2$CoRuO$_6$ (A = Pr, Nd, Sm and Y), where the magnetic data of Y$_2$CoRuO$_6$ is obtained from reference~\cite{y2coruo6}. The moments at 5 K ($M_{5K}$), on applying  50 kOe of field,  is found to be 2.3, 2.9, 1.4 and 0.8 $\mu_B$/f.u. on the virgin line of the $M-H$ curve for Pr, Nd, Sm and Y compounds. Y does not carry any moment, while the total angular momentum $J$ = 4, 9/2 and 5/2 for Pr$^{3+}$, Nd$^{3+}$ and Sm$^{3+}$ states respectively. The  magnetic moments of these three  ions are 3.58, 3.62 and 0.84 $\mu_B$ respectively. Clearly, the variation of $M_{5K}$ corresponds well with the variation of  rare-earth moment. This signifies that the A site rare-earth moment remains in ordered state at 5 K. It is important to perform a neutron diffraction study to ascertain the true magnetic structure of these compounds.          

\par
In summary we have studied the structural, magnetic as well as transport properties of the  double perovskites La$_2$CoRuO$_6$, Pr$_2$CoRuO$_6$, Nd$_2$CoRuO$_6$, Sm$_2$CoRuO$_6$. We observe a systematic change in the magnetic ground state as La is replaced by Pr, Nd and Sm. This matches well with the case of Fe-Os and Co-Os based double perovskites, where lattice distortion tunes the  strength of the magnetic interactions  in different exchange pathways. 

\section{Acknowledgment}
The work is supported by the financial grant from DST-SERB project (EMR/2017/001058). MD would like to thank CSIR, India for her research fellowship, while PD thanks DST-SERB for his NPDF fellowship (PDF/2017/001061). 

%%%%%%%%%%%%%%%%%%%%%%%%%%%%%%%%%%%%%%%%%%%%%%%%%%%-Reference-%%%%%%%%%%%%%%%%%%%%%%%%%%%%%%%%%%%%%%%%%%%%%%%%%%%%%%%%%%%%%%%%

\end{document}